\documentclass[preprint,pre,superscriptaddress,floatfix,aps]{revtex4}
\newcommand{\beq}{\begin{equation}}
\newcommand{\eeq}{\end{equation}}
\newcommand{\beqa}{\begin{eqnarray}}
\newcommand{\eeqa}{\end{eqnarray}}

\usepackage{graphicx}
\usepackage{dcolumn}
\usepackage{bm}
\usepackage{longtable}
\usepackage{amsmath}
\usepackage{amssymb}
\usepackage{color}

\begin{document}
\title{Crystalline free energies of micelles of diblock copolymer solutions}
\author{Giuseppe D'Adamo}
\affiliation{Department of Physics, University of L'Aquila, Via
Vetoio 10, I-67010 L'Aquila, Italy}
\author{Carlo Pierleoni}
\affiliation{CNISM and Department of Physics, University of L'Aquila, Via
Vetoio 10, I-67010 L'Aquila, Italy}
\email{carlo.pierleoni@aquila.infn.it}


\begin{abstract}
We report a characterization of the relative stability and structural behavior of various micellar crystals of an athermal model of AB-diblock copolymers in solution.  We adopt a previously developed coarse-graining representation of the chains which maps each copolymer on a soft dumbbell. Thanks to this strong reduction of degrees of freedom, we are able to investigate large aggregated systems, and for a specific length ratio of the blocks $f=M_A/(M_A+M_B)=0.6$, to locate the order-disorder transition of the system of micelles. Above the transition, mechanical and thermal properties are found to depend on the number of particles per lattice site in the simulation box, and the application of a recent methodology for multiple occupancy crystals (B.M. Mladek et al., Phys. Rev. Lett.  \textbf{99}, 235702 (2007)) is necessary to correctly define the equilibrium state. Within this scheme we have  performed free energy calculations at two reduced density $\rho/\rho*=4,5$ and for several cubic structures as FCC,BCC,A15.  At both densities, the BCC symmetry is found to correspond to the minimum of the unconstrained free energy, that is to the stable symmetry among the few considered, while the A15 structure is almost degenerate, indicating that the present system prefers to crystallize in less packed structures. At $\rho/\rho*=4$ close to melting, the Lindemann ratio is fairly high ($\sim 0.29$) and the concentration of vacancies is roughly $6\%$.  At $\rho/\rho*=5$ the mechanical stability of the stable BCC structure increases and the concentration of vacancies accordingly decreases. The ratio of the corona layer thickness to the core radius is found to be in good agreement with experimental data for poly(styrene-b-isoprene)(22-12) in isoprene selective solvent which is  also reported to crystallize in the BCC structure. \end{abstract}

\pacs{PACS:  }
\maketitle

\section{Introduction}
 \noindent\label{sec:Introduction,1}
A recurrent phenomenology in soft matter is the self-assembly of its basic constituents into supramolecular aggregates of various shapes and morphologies. This property, ubiquitous in materials of biological and technological interest, greatly enriches the phase behavior of these systems with the possibility of order-disorder transition of aggregates on mesoscopic time and space scales. A remarkable example in this sense is represented by diblock copolymers both in melts\cite{Leibler80,Bates91} and solution\cite{Lodge05,Hamley05}. The occurrence of micro-phase separations into a large number of spatially patterned arrangements (lamellar, gyroid) or clusters of different shape and size (cylinders, micelles) in diblock copolymer solutions has been experimentally characterized\cite{Lodge05,Hamley05,McConnell93,BangLodge08}. For clustered conditions, it has also been observed the occurrence of a transition between a disordered phase of dispersed clusters and a phase with long range spatial order in hexagonal (for cylindrical aggregates) or cubic (for spherical clusters) structures (ODT). Moreover thermotropic and lyotropic transitions between cubic crystals of different structure have been reported \cite{Bang02,Lodge02,Bang04,Lai} and theoretically analyzed by mean field theoretical arguments\cite{Grason,Zhulina05,Suo}.

Thanks to the  number of available control parameters, such as temperature, concentration of the copolymer, solvent selectivity with respect to the two  blocks and the relative block lengths, these systems are in principle very versatile and could be designed to obtain a specific phase behavior. In practice however this program requires a microscopic understanding of the entropic-enthalpic competition behind the self-assembling and the ODT which is  still missing. This complex picture requires a considerable effort both in theoretical  and molecular simulation approaches, and while in  the melt regime the low osmotic compressibility allows to successfully apply mean field theory \cite{Leibler80,BatesFredrickson90}, in solutions the possibility to make ab initio predictions is more difficult\cite{Cavallo06}. Even for implicit solvent models the application of the common molecular simulation strategies is limited by the huge number of degrees of freedom in the system (many long chains) and the wide range of time and length scales involved\cite{Milchev01,Cavallo06}. 
In order to describe the phase behavior of large aggregated systems it is tempting to use a coarse graining scheme which allows to reduce the number of degrees of freedom by mapping a number of monomers of each chain onto soft blobs. This strategy goes back to a seminal work of Louis et al. \cite{LouisBolhuisHansenMeijer00} which mapped a solution of homopolymers onto a fluid of soft colloids interacting through, density dependent, soft core, pair potentials obtained from the observed structure through liquid state theory methods. Working with density dependent potentials is however cumbersome\cite{Louis02}, in particular for inhomogeneous systems, and can be avoided by switching to the so called ``multi-blob" representation, in which the degree of coarse graining (the number of monomer mapped onto a single blob) is tuned on the local density. This scheme has been applied to a semi-dilute solution of homopolymers in good solvent\cite{PierleoniCaponeHansen07} and very recently an extension to diblock copolymer solutions has been proposed\cite{CaponeHansenColuzza10}.

Within this strategy, the coarsest grained representation of diblock copolymers maps each block of a chain onto its center of mass reducing a long polymer to a soft dumbbell. This is the minimal model of diblock copolymers and it is accurate at high dilution only. Its extension to finite concentration (i.e. determining density dependent potentials as in the case of homopolymers) is not trivial since the theoretical connection between structure and pair interactions at finite density is missing. Nonetheless the minimal model with zero density interactions, extensively investigated by Monte Carlo (MC) simulations well beyond the overlap concentration \cite{Addison05,Pierleoni06,Capone09}, exhibits the self-assembling phenomena and a number of ODT transitions in qualitative agreement with the experimental phenomenology of diblock copolymer solutions. 
This model has been studied for various solvent selectivity and over a wide range of densities and block lengths ratios\cite{Addison05,Capone09}. It has been found that the ISI model, in which each block of the copolymer is in an ideal (theta) solvent while pairs of monomers of the two blocks repel each other (mutually-avoiding), has a microphase separation towards a lamellar phase beyond a given concentration\cite{Addison05}, whereas a ISS model, in which one block is in theta solvent while the other is in a good solvent and pairs of monomers of  differents block are mutually-avoiding, naturally form spherical micelles beyond a CMC density \cite{Pierleoni06,Capone09}. Moreover in the ISS model, an order-disorder transition (ODT) between a liquid and a crystalline phase of micelles is spontaneously observed to occur for high enough concentration. The same qualitative picture emerges from a ``multi-blob" investigation of the ISS model \cite{CaponeHansenColuzza10} with the addition of an observed cylindrical phase. Because of the large aggregation number ($\sim100$), the number of micelles in a simulation sample is limited and the structure of the crystals that spontaneously form is influenced by the constraints of the simulation and by the slow relaxation of the system. In order to obtain detailed predictions about the relative stability of disordered and ordered phases of different symmetries, free energy calculations are necessary. 

The computational approach to the free energy of these complex systems is significantly more difficult than the usual methods employed in simple non-assembling systems. 
On the one hand, the soft interactions  between blobs makes the use of Widom insertion method highly efficient even at high density and even in the crystal phase (the phase in which micelles are ordered according to a lattice). On the other hand, the softness of the potential produces floppy and polydisperse aggregates which pose a fundamental problem to the calculation of the free energy of crystalline phases. The problem is related to how correctly to define the equilibrium state of the system with broken symmetry, in particular to determine the equilibrium number of particles per lattice site for a given density (and temperature in the general case). This problem does not arise in crystals with single-particle occupancy (one particle per lattice site) but has been shown to be relevant  for multiple-particle occupancy crystals \cite{Mladek1,Mladek2}. 
For the minimal dumbbell model of dibock copolymers beyond the ODT, the size distribution of the micelles is found to depend significantly on the number of particles in the sample and a given crystal structure with fixed number of lattice sites can comfortably accommodate a wide range of particle numbers, by adjusting the size of the aggregates, before loosing its structure. In a recent paper on multiple-occupancy crystals, using ideas from the thermodynamic theory of crystals with vacancies \cite{Swope}, Mladek et al.\cite{Mladek1,Mladek2} devised a method to define the equilibrium number of particles per lattice site and a strategy to compute the free energy of the equilibrium system. In the present paper we apply the same formalism and method to compute the free energy of cubic crystals of spherical micelles of diblock copolymers in solution. Since the method is quite demanding in terms of computer time, we adopt the dumbbells minimal model of diblock copolymers despite the fact that its use at finite density is not fully justified. Future work will consider the more accurate multiblob representation.

Finally, we should mention a previous attempt to address the relative stability of crystals of different symmetries of diblock copolymers micelles \cite{Molina09}. In this approach micelles are mapped onto point particles interacting through density dependent effective pair potentials obtained by inverting the micelle-micelle radial distribution function. 
This work suggests that in the range $0.4\leq f \leq 0.6$ the OD transition density decreases upon increasing $f$, and that the A15 structure\cite{WeairePhelan94}, already proposed as the stable structure for spherical micelles in linear block copolymer melts \cite{Kamien,Grason03}, in start block copolymer melts \cite{GrasonKamien04,Grason05} and experimentally observed for dendrimic clusters \cite{Hudson}, is the most stable structure among the few explored, namely BCC, FCC, diamond, SC, A15. This is somehow in contradiction with the expectation that the favored structure should depend on $f=M_A/(M_A+M_B)$, with micelles with smaller coronae (crew-cut) preferring close packed structures such as FCC or HCP, while micelles with larger coronae (hairy) prefer less packed structures such as BCC and A15 \cite{McConnell93,Grason,BangLodge08}. However this effective micelle model differs from the original system of dumbbell micelles in several important aspects. At the dumbbell level, the micelles are highly polydisperse presenting a gaussian-like aggregation number distribution peaked around 100 dumbbells with a width of roughly 40 dumbbells\cite{Pierleoni06,Capone09}. The polydispersity is lost in mapping micelles onto point particles. Moreover, the number of aggregates in the system at the dumbbell level of description is found to fluctuate producing a crystal with a finite concentration of vacancies, this might be mainly due to the floppy character of the aggregates. Finally, even at densities beyond the observed ODT, in the system of dumbbells some molecules remain dispersed in the volume between the clusters, establishing a ``chemical" equilibrium between the aggregated and non aggregated molecules. Again this phenomenon is lost when mapping micelles onto point particles, although it is implicitly taken into account in the effective interactions.  Because of such important differences the prediction of the ``effective" micelles system can be only considered as qualitative and a definitive assessment should be provided by a direct calculation of crystal free energies at the dumbbell level of description. 

The paper is organized as follows. In section \ref{sec:CoarseGrainedModel} we briefly review our model which has been already presented in previous works\cite{Pierleoni06,Capone09}. In the following sections \ref{sec:Thermo} and \ref{sec:ComputationalAspects} we review the thermodynamic theory of cluster crystals and the method to compute the crystal free energy. We also describe specific aspects related to our present system. In section \ref{sec:liquid} we report results for the disordered phase and the approach to the spontaneous OD transition which were not completely discussed in previous publications. In the following section we discuss our results of the free energy calculations of several crystalline structures at two different densities. Finally in section \ref{sec:Conclusion} we draw our conclusions and perspectives. Appendices I and II are devoted to the calculations of the free energy of the reference system, and to details about the application of the Gauss-Legendre method to perform the coupling constant integration, respectively.

\section{Coarse-grained Model} \label{sec:CoarseGrainedModel}
Let us consider a system of N copolymer chains, each formed by two blocks A and B of $M_A$ and $M_B$ monomers of different chemical species, in a volume V at fixed chain density $\rho=N/V$. We consider an implicit solvent model in which solvent degrees of freedom are absent and the solvent selectivity is represented by effective monomer-monomer interactions specific for each different pair of particles. We consider here the ISS athermal model representing theta solvent conditions for monomers of the A block and good solvent conditions for monomer of the block B. Moreover the interaction between monomers of A and B blocks is mutually-avoiding. The phase diagram of this model is fully determined by entropic effects and the only independent variables are the chain density $\rho$ and the relative size of the blocks $f=M_A/(M_A+M_B)$. As usual in polymer solutions \cite{Doi}, the density is better expressed in terms of the reduced density $\rho/\rho^*$ where $\rho^*=3/(4\pi R_g^3)$ is  the overlapping density and $R_g$ is the radius of gyration of an isolated chain. Direct simulation of this full monomer model in the scaling regime (i.e. for long enough chains) is challenging because of the large number of degrees of freedom in the system ($N\times M$) and the wide range of relaxation times involved. 

The simplest coarse grained representation of this model is obtained by mapping each block on its centre of mass therefore reducing each chain to a simple dumbbell \cite{Addison05,Pierleoni06,Capone09}. The interaction potentials between the two blobs of the same dumbbell $\gamma_{AB}(r)$ and between blobs of different dumbbells, $v_{AA}(r), v_{AB}(r), v_{BB}(r)$ can in principle be obtained by inverting the structure obtained in a full monomer simulation. This is a well known procedure for single and multicomponent systems of homopolymers\cite{Addison051}. For diblock copolymers the connectivity constraint makes the problem well defined only at zero density, i.e. for two isolated dumbbells\cite{Addison05}. In this case the block center of mass radial distribution functions, obtained by Monte Carlo simulation of a system of two interacting chains at the full monomer level, can be inverted to get the pair interaction potentials between the blobs\cite{Addison05}. However it is not known how to consistently extend this procedure to finite density. Therefore to study the interesting thermodynamic behavior of this system at finite density we must rely on the approximation of using density independent potentials identical to the ones extracted at zero density. As already observed \cite{Pierleoni06}, these potentials do not differ much from the potentials obtained for mixtures of untethered chains at zero density. In the absence of the tethering constraint between A and B blocks we have a mixture of two chemically different homopolymers in the presence of solvent of different quality. In this case the multicomponent HNC closure could be exploited to invert the pair structure and therefore extract pair potentials between blobs (center of mass of the homopolymers) even at finite density. However the physical behavior of this mixture at finite density is radically different from the behavior of the diblock copolymer solutions. Indeed in the mixture a macroscopic phase separation occurs at large enough density while in the copolymer system only microphase separations, such as micellization, occurs. It is therefore clear that the analogy between the two sets of potentials cannot be extended to finite density. 

As in previous investigations \cite{Pierleoni06,Capone09} we adopt here the potentials for a mixture of A and B homopolymers at zero concentration.We assume that the A chains are ideal (theta solvent) and the B chains are self-avoding (good solvent) and that the interaction between monomers of the A and the B chains are mutually-avoiding. At zero density the potentials between the centres of mass of any pair of such polymers $v_{\alpha\beta}(r), (\alpha, \beta=A,B)$ are the potentials of mean force   
\beq
\label{eq:potential}
v_{\alpha\beta}(r)=-k_BT \log g_{\alpha\beta}(r)
\eeq
where $g_{\alpha\beta}(r)$, the centre of mass radial distribution functions, are obtained by MC simulations of a systems of two chains on lattice at the full monomer level.
Morever, in order to restore the connectivity constraint in the diblock copolymer system, the tethering potential $\gamma_{AB}(r)$ between the A and B blocks of the same copolymer, is consistently obtained at zero density from the radial distribution function between the centre of mass of the two blocks $s_{AB}(r)$ obtained by MC simulation of a single diblock chain at the full monomer level
\beq
\gamma_{AB}(r)=-k_BT\log s_{AB}(r)
\eeq

\begin{figure}
\center
\includegraphics[scale=0.32]{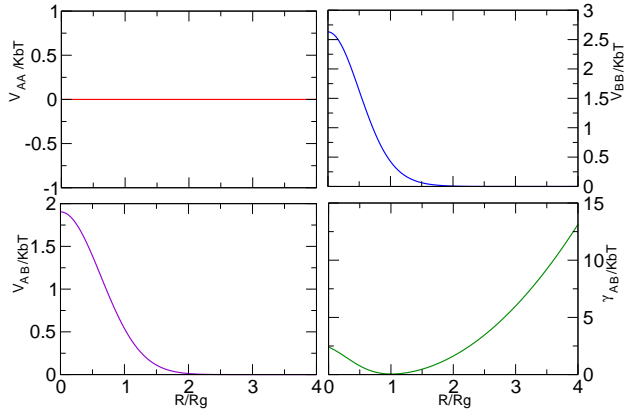}
\caption{f=0.6: effective pair potentials $v_{AA},v_{AB}, v_{BB},\gamma_{AB}$, as a function of the distance expressed in gyration radius units $R_{g}$.}
\label{Fig:potentials}
\end{figure}

The potentials are reported in Fig. \ref{Fig:potentials}.
The effective intermolecular potential $v_{AA}(r)=0$ while $v_{AB}(r)$ and $v_{BB}(r)$ are soft and roughly gaussian with a characteristic length comparable to the copolymers gyration radius, and an amplitude of roughly $2k_{b}T$ at full overlap.
We make the assumption that the zero density potentials can be used at finite density. This transferability assumption is supported by the evidence of a weak density dependence of the effective potentials for homopolymers\cite{LouisBolhuisHansenMeijer00}.


In previous works\cite{Pierleoni06,Capone09} the thermodynamic behavior of this model has been explored. Beyond some value of the reduced density which depends on $f$, the dumbbells self-assemble into spherical micelles. At the same time thermodynamic properties such as the excess pressure and the excess chemical potential exhibit a change of behavior providing a macroscopic signature of the CMC and of the microphase separation. Upon further increase of the reduced density a ODT is observed with a spontaneous breaking of the translation symmetry and an ordering of micelles according to some crystalline structure. The precise location of the CMC and of the ODT is obviously influenced by the zero density potential approximation, but the qualitative picture is confirmed by the recent multi-blob exploration\cite{CaponeHansenColuzza10}.

To complete the thermodynamic characterization of the dumbbell model it would be desirable to acquire knowledge about the stable crystalline structures that one should expect beyond the ODT. Indeed different crystalline phases under different thermodynamic conditions have been reported in experiments \cite{Lodge05} and a number of experimental studies about solid-liquid and solid-solid phase transition in diblock-copolymer micelles have appeared \cite{McConnell93, Bang02, Bang04,Lodge05}.  It is tempting to explore the capability of our simple model to qualitatively reproduce such transitions. As mentioned above however free energy calculations for micellar crystals present
some inherent difficulty related to the definition of the true equilibrium state in a simulation system with Periodic Boundary Conditions. Naively one could think that the free energy of a system of micelles for a given $f$ and at given reduced density $\rho/\rho^*$, ordered along the nodes of a specific crystalline lattice, can be simply obtained by computing the single molecule chemical potential by the Widom insertion method, which remains very efficient even at relatively high density, and the pressure by the virial expression. However our floppy micelles are able to accommodate a variable number of dumbbells and  in a periodic system at fixed density and fixed number of micelles (each one at a different lattice node) the number of molecules and the corresponding volume can vary in a finite range before the crystalline state becomes dynamically unstable. This anomaly is a clear signature of the presence of an additional hidden variable. As explained in the next section this variable is the number of lattice nodes which, in a system with Periodic Boundary Conditions (PBC), plays the role of a thermodynamic constraint. An analogous situation has been recently found in the multiple occupancy crystals, i.e. systems which in the crystalline phase are able to accommodate an increasing number of particles per lattice site upon increasing the external pressure \cite{Mladek2,Mladek1}. 

\section{Thermodynamics of cluster crystals} \label{sec:Thermo}
Our presentation of the theory and of the strategy to compute the crystal free energies follows closely the original work of Swope and Anderson \cite{Swope}  and a recent implementation by Mladek et al. \cite{Mladek1,Mladek2}. In the study of single occupancy crystals, it is common practice to associate the equilibrium state of the system with the defect-free state, except when one is interested in the properties of the defects. Indeed in most cases the equilibrium concentration of vacancies or defects is so low that this association is quantitatively correct \cite{Frenkel}. For multiple occupancy or cluster crystals, characteristic of soft matter physics, this is no longer true because a number of simple constituents  can share the same site.  
In real systems through surface adjustments and the spontaneous occurrence of defects  it is possible to reach the equilibrium minimum  free energy, that optimizes the ratio $N/N_{s}$ between the number of lattice sites $N_{s}$, and the number of particles $N$. Conversely in simulations, periodic boundary conditions impose a constraint on the number of  lattice sites  $N_{s}$, and at fixed number of molecules $N$, volume $V$ and temperature $T$  this prevents to reach the equilibrium state with the corresponding equilibrium ratio  $N/N_{s}$ . In thermodynamic terms, the free energy cost to add or remove a lattice site, that we denote with $ \mu_{c}$, is too high to observe adjustments on typical simulation time scale. The free energy of the constrained system with fixed $N_{s}$ is determined not only by the conjugate couples $(S,T),(\Pi,V),(\mu,N)$, where $S$ is the entropy, $\Pi$ the osmotic pressure and $\mu$ the chemical potential, but also by the site chemical potential $\mu_{c}$  and its conjugate variable $N_{s}$. In the constrained system,  an infinitesimal variation of $F$ can be written as:
\beq
\label{eq:InfyFree}
dF(\mu_{c})=-S(\mu_{c})dT-\Pi(\mu_{c}) dV+\mu(\mu_{c}) dN+\mu_{c} dN_{s}
\eeq
The expressions for the main thermodynamic constrained properties follow from the previous equation:
\beqa
\label{eq:ConstrainedQuanty}
\Pi (\mu_{c})=-\left(\frac{\partial F}{\partial V}\right)_{N_{s}NT} \nonumber
\\
\mu(\mu_{c})=\left(\frac{\partial F}{\partial N}\right)_{N_{s}VT}
\eeqa
The thermodynamic derivates, like the pressure and chemical potential, are performed keeping $N_{s}$ fixed, and this leads to consider  $\mu_{c}$ as a constraint  to the other properties. However $N_{s}$, as well as $\mu_{c}$, is an unphysical variable and the results of the constrained system will be representative of the unconstrained situation only when site chemical potential $\mu_c$ vanishes. Incidentally this condition restores the extensivity of the Gibbs free energy with the number of molecules in the system.
 \beqa
 \label{eq:Gibbs}
G=F+PV=\mu N+\mu_{c}N_{s} \rightarrow \mu_{c}=0
\eeqa
This condition is equivalent to say that the free energy cost to add or remove a site at equilibrium is zero and corresponds to the equilibrium  value of the constraint.  From a computational point of view the locus of points $\mu_{c}(\rho,T) = 0$, is obtained through canonical simulations at fixed densities $\rho$ and temperatures $T$, varying  both $N$ and $V$ to find the specific values $\bar{N},\bar{V}$ for which:
\beq
\label{eq:equilibrium}
\mu_{c}(\bar{N},\bar{V},T)|_{\rho=\frac{\bar{N}}{\bar{V}}}=0=
\frac{ 
[F(\bar{N},\bar{V},T)+\Pi(\bar{N},\bar{V},T) \bar{V}-\mu \bar{N}]
       } 
       {N_{s}}
\eeq
\indent In order to determine the values $(\bar{N},\bar{V})$ three independent measurements of $F, \Pi, \mu$ are necessary. Following this formalism, we can recover predictions on the equilibrium unconstrained system from a constrained simulation imposing eq. (\ref{eq:equilibrium}):
\beqa
\Pi(\bar{N},\bar{V},T) = \Pi(\mu_{c} = 0) = \Pi(\bar{N},\bar{V},T,N_{s}^{eq}(\bar{N},\bar{V},T))	 \nonumber
\\
\mu(\bar{N},\bar{V},T) = \mu(\mu_{c} = 0) = \mu(\bar{N},\bar{V},T,N_{s}^{eq}(\bar{N},\bar{V},T))	
\eeqa

\section{Computational aspects} \label{sec:ComputationalAspects}
\begin{figure}
\includegraphics[scale=0.35]{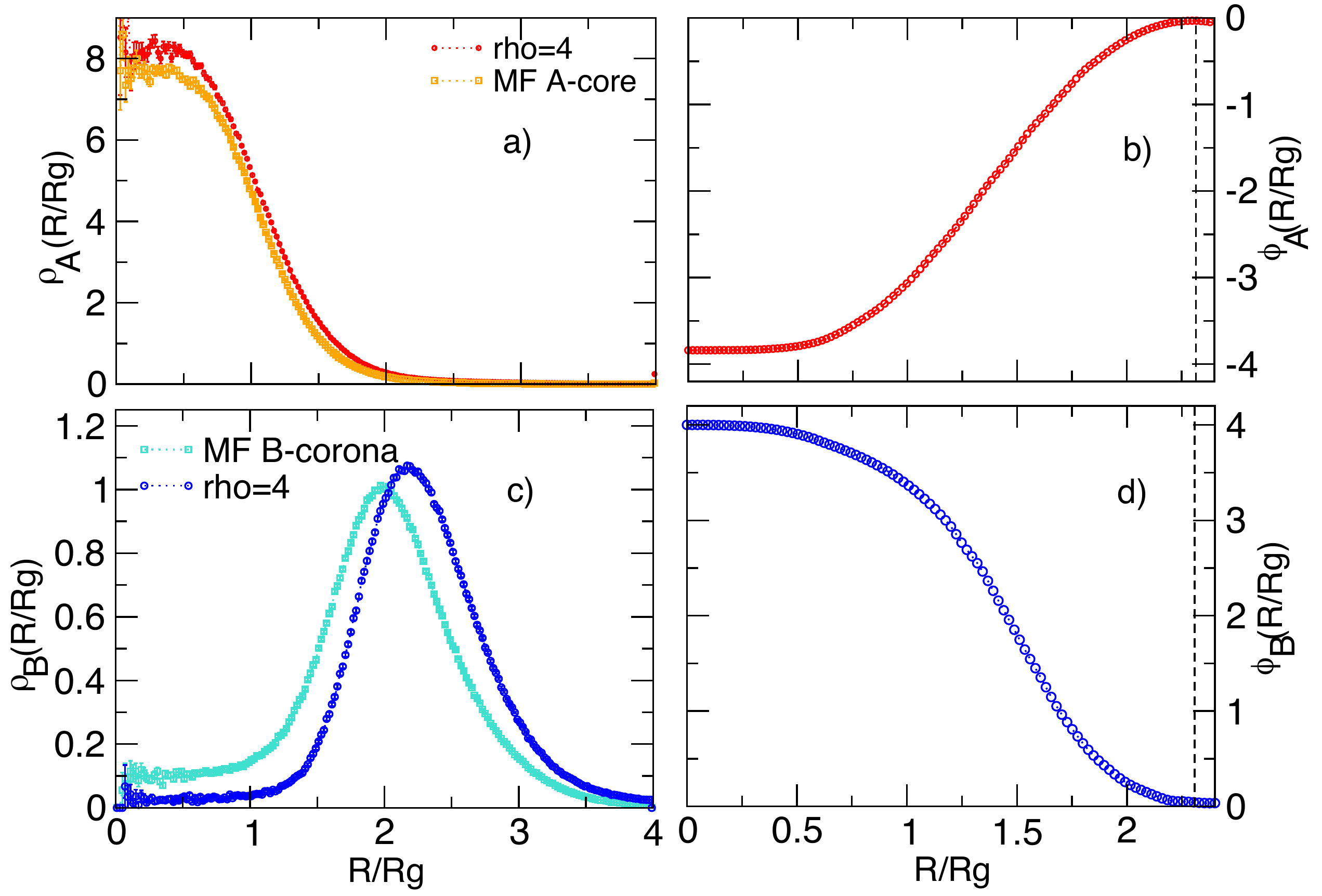}
\caption{Radial density of core $\rho_A$ and corona $\rho_B$ (left hand), and effective external field $\phi_A$,$\phi_B$ (right hand) as function of the reduced distance $R/Rg$. a),c) density profiles for the real and the effective system for $\rho/\rho^*=4$; b),d) corresponding external fields, the vertical dashed line represent the potential cutoff $r_c=2.3$. }
\label{Fig:mfpotentials}
\end{figure}

For each thermodynamic state $(\rho,T)$ it is necessary to perform several simulations in the constrained ensemble $N,V,N_{s},T$,  varying $N/N_s$ and correspondingly $V/N_s$ and computing ${F,\mu,\Pi}$ until eq.(\ref{eq:equilibrium}) is verified within the computational uncertainties. The estimate of the free energy, the osmotic pressure and the chemical potential, is achieved  through a thermodynamic integration, the virial expression and the Widom insertion/removal method respectively\cite{FrenkelSmit}. The possibility to use the Widom insertion method in a dense phase is guaranteed by the softness of the interaction potentials. The implementation of Kirkwood coupling constant integration method \cite{Kirkwood} is significantly different from the usual Frenkel Ladd method\cite{FrenkelSmit}. In the micellar crystals polymers can migrate from one aggregate to the next and therefore a molecule cannot be assigned univocally to a specific lattice site. To preserve such a feature along the coupling constant integration path, it is convenient to assume as reference system an ideal gas of dumbbells in an external field $\Phi_{ext}$ that mimics the desired lattice geometry. The form of the external potential must allow the single molecule to jump from one Wigner-Size cell to the next as it is observed in the physical system.  This can be achieved by summing over lattice sites, a dumbbell-site central pair interaction which vanishes at the edge of the WS cell. To compute the free energy difference between the original and the reference system, we employ a linear parametrization:
\beqa
\label{eq:interaction}
U(\lambda)=U_{intra}+(1-\lambda)U_{inter}+\lambda\Phi_{ext}
\eeqa
where $U_{intra}=\sum_{i}^{N}\gamma_{AB}(r_{i})$ is the intradumbbell potential energy, $U_{inter}=\sum_{i<j}^{N}(v_{AB}+v_{BB})_{ij}$ is the interdumbell potential energy and $\Phi_{ext}=\Phi_{A}+\Phi_{B}$ is the energy of the external field acting on the A and the B blocks. 
The free energy is then obtained as \cite{Kirkwood}:
\beq
\label{eq:free}
F(\lambda=0)=F({\lambda=1})+\int_{1}^{0}\langle \Phi_{ext}-U_{inter} \rangle_{\lambda}~d\lambda
\eeq
where $\lambda=1$ corresponds to the reference system. The dumbbell-site interaction is in principle arbitrary but for reason of efficiency it must be chosen to better reproduce some property of the original system. In the present case we have taken external fields acting on  both A and the B blocks in such a way to roughly reproduce the density profiles, $\rho_A(r), \rho_B(r)$ within a micelle as obtained in a crystalline state simulation of the orginal system of interacting dumbbells. Strictly speaking, the inversion procedure from density profiles to single-particle interaction, requires the solution of two coupled integral equations due to the intradumbbell correlation. We have used a simplified approach based on simple barometric relations:

\beq
\psi_{\alpha}(r_{\alpha})=-\lambda^{-3}_{\alpha}\log\langle \rho_{\alpha}(r_{\alpha})\rangle+\Theta_{\alpha} \\ 
\eeq
where $\alpha=A,B$ and $\Theta_{\alpha}$ and $\lambda_{\alpha}$ are additive and dimensional constants respectively.
In order to form a smooth function through the three dimensional space, it is important that each potential and its first derivative vanish at the edge of the Wigner-Size cell $r_c$.
To this aim we regularize and cut the above forms as 
\beq
\phi_{\alpha}(r)=\left\{ 
\begin{matrix}
&\psi_{\alpha}(r)+\psi_{\alpha}(2r_c-r)-2\psi_{\alpha}(r_c) \qquad\qquad &r\leq r_c \\
&0 \qquad\qquad\qquad\qquad\qquad  &r > r_c
\end{matrix}\right.
\eeq
These potentials, reported in fig. \ref{Fig:mfpotentials}, represent the effective interactions between one lattice site and a block of a generic dumbbell.
The external field to be used in our coupling constant integration is built by summing all contributions from different nodes of a lattice of the chosen symmetry $\{R_{m}\}, m=1,...,N_s$:
\beq
\label{eq:effectivePhi}
\Phi_{\alpha}(r_{\alpha})=\sum_{m=1}^{N_{s}} \phi_{\alpha}(r_{\alpha}-R_{m}) \qquad\qquad\qquad \alpha=A,B
\eeq
The free energy of the reference system, i.e. a system of non interacting dumbbells in the lattice field of eq.(\ref{eq:effectivePhi}) cannot be computed analitycally, but can be recast in the form of a statistical average and computed by MC simulation, as detailed in Appendix I. 
The numerical estimate of the integral over $\lambda$ in eq. (\ref{eq:free}) is obtained through the Gauss-Legendre quadrature scheme as detailed in Appendix II. 

\section{Dispersion of micelles and OD transition}\label{sec:liquid}
In the present work we limit our investigation to a single value of $f=0.6$. The main reason is the large computational demand of the present technique. The other, more fundamental, reason is the limitation of the present dumbbell model, based on the zero density potential approximation, to investigate the intermediate density regime. Therefore this work is mainly meant as a methodological development to prepare the stage for studying the more accurate multiblob representation of diblock copolymer solutions \cite{CaponeHansenColuzza10}.

\begin{figure}
\center
\includegraphics[scale=0.33]{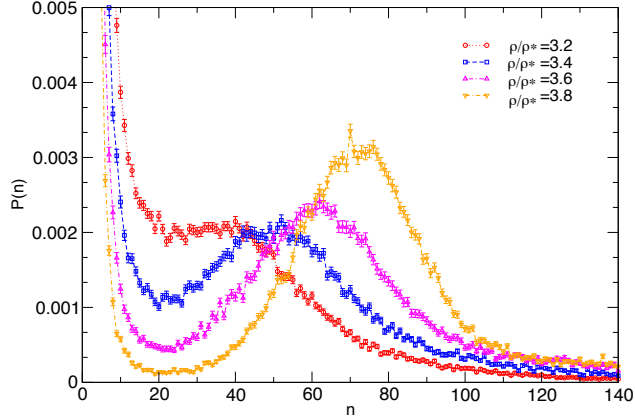}
\caption{Probability of a cluster of n molecules P(n), for different densities above the CMC.}
\label{fig:pn}
\end{figure}

\begin{figure}
\center
\includegraphics[scale=0.33]{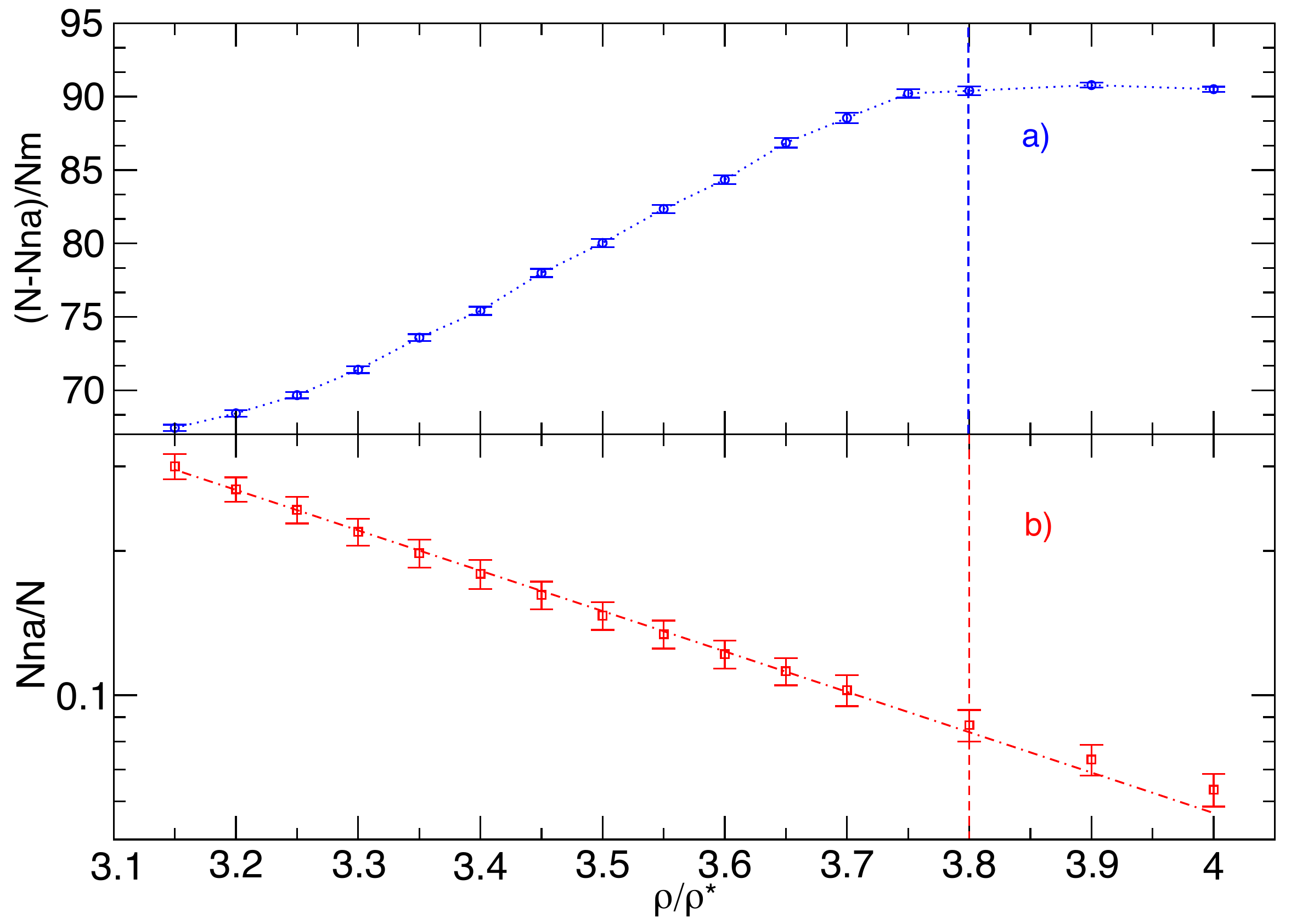}
\caption{Panel a) estimate of the mean aggregation number of micelles as function of the reduced density; panel b) number of dispersed dumbbells as function of the reduced density on a semi-log scale, the dash-dotted line represents an exponential fit to the data.  The dashed vertical lines indicate the estimated freezing density.}
\label{fig:Nmic}
\end{figure}

\begin{figure}
\center
\includegraphics[scale=0.33]{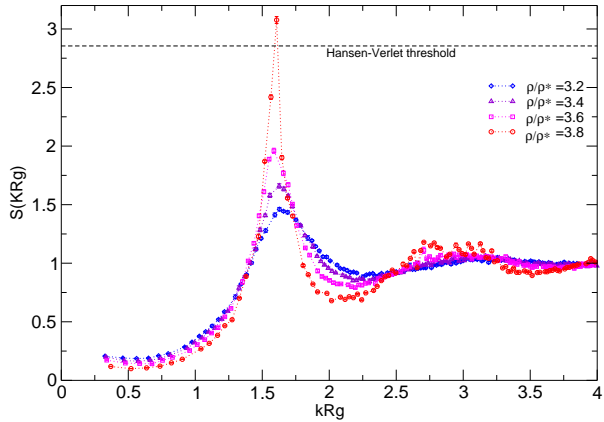}
\caption{Micelle-micelle structure factors $S(k)$ as function of the reduced density above the CMC. For $\rho/\rho^*=3.8$, the first peak exceeds the Hansen-Verlet threshold (dashed line).}
\label{fig:SKmic}
\end{figure}

We have performed MC simulations within the canonical ensemble, at various densities in the range $\rho\in[3.0,5.0]$. In a previous work \cite{Capone09} the density of the onset of ordering was observed to decrease with increasing f, and we have chosen f=0.6 in order to explore the solid region without stretching too much the transferability hypothesis on the potentials. Clusters structure is characterized through the study of the distribution function $P(n)$ that describes the probability of finding a cluster of size $n$.  Above the CMC, this quantity shows a clear bimodal behavior with an exponential peak at the origin, signature of dispersed dumbbells, and a gaussian-like peak at a finite values of $n$ signaling the presence of well defined clusters (see Fig.\ref{fig:pn}). 
The width of the cluster peak decreases with increasing density, meaning a smaller polydispersity and more stable aggregates at higher density. Note that $P(n)$ has a minimum at $n\simeq 20$ so that we consider an aggregate to be a cluster only when it includes more than 20 molecules.  
In Fig. \ref{fig:Nmic} we report the average cluster size, computed as the ratio between the total number of molecules belonging to clusters $\langle N-N_{na}\rangle$ and the average number of detected aggregates $\langle N_m\rangle$ (panel a), and the fraction of non-aggregated dumbbells $\langle N_{na} \rangle /N$ (panel b) as a function of the reduced density. Note that results at different densities are obtained with different system sizes and the smoothness of the behavior confirms the uniqueness of the equilibrium state for the system in the disordered phase, in contrast to the case of the ordered state (see below). We observe a roughly linear increase of the cluster size $\langle N-N_{na}\rangle/\langle N_{m}\rangle$ up to the value $\rho/\rho^*=3.8$ above which the slope decreases considerably. This change of behavior signals the freezing transition for the micelles.
The same indication can be obtained from the structure factor between the centres of mass of micelles 
\beq
S(k)=\langle \frac{1}{N_{cluster}}\sum_{i\ne j}^{N_{cluster}}e^{-i\mathbf{k}\cdot(\mathbf{R}_{cm}^{i}-\mathbf{R}_{cm}^{j}) }\rangle
\eeq
reported in fig. \ref{fig:SKmic}, where we see that at $\rho/\rho^*=3.8$ the amplitude of the first peak exceeds the Hansen-Verlet threshold value for freezing \cite{HansenVerl}. This criterion, proposed for simple fluids, it is found to hold also in the present case of micellar fluids. Indeed beyond $\rho/\rho^*=3.8$ we observe a slowing down of the micellar motion due to the sampling, despite the fact that the efficiency of the Monte Carlo sampling of single molecular moves is not reduced.
Panel (b) of fig. \ref{fig:Nmic} reports an exponential decrease of the number of non-aggregated molecules with density
with only a marginal deviation above freezing. 

\begin{figure}
\center
\includegraphics[scale=0.33]{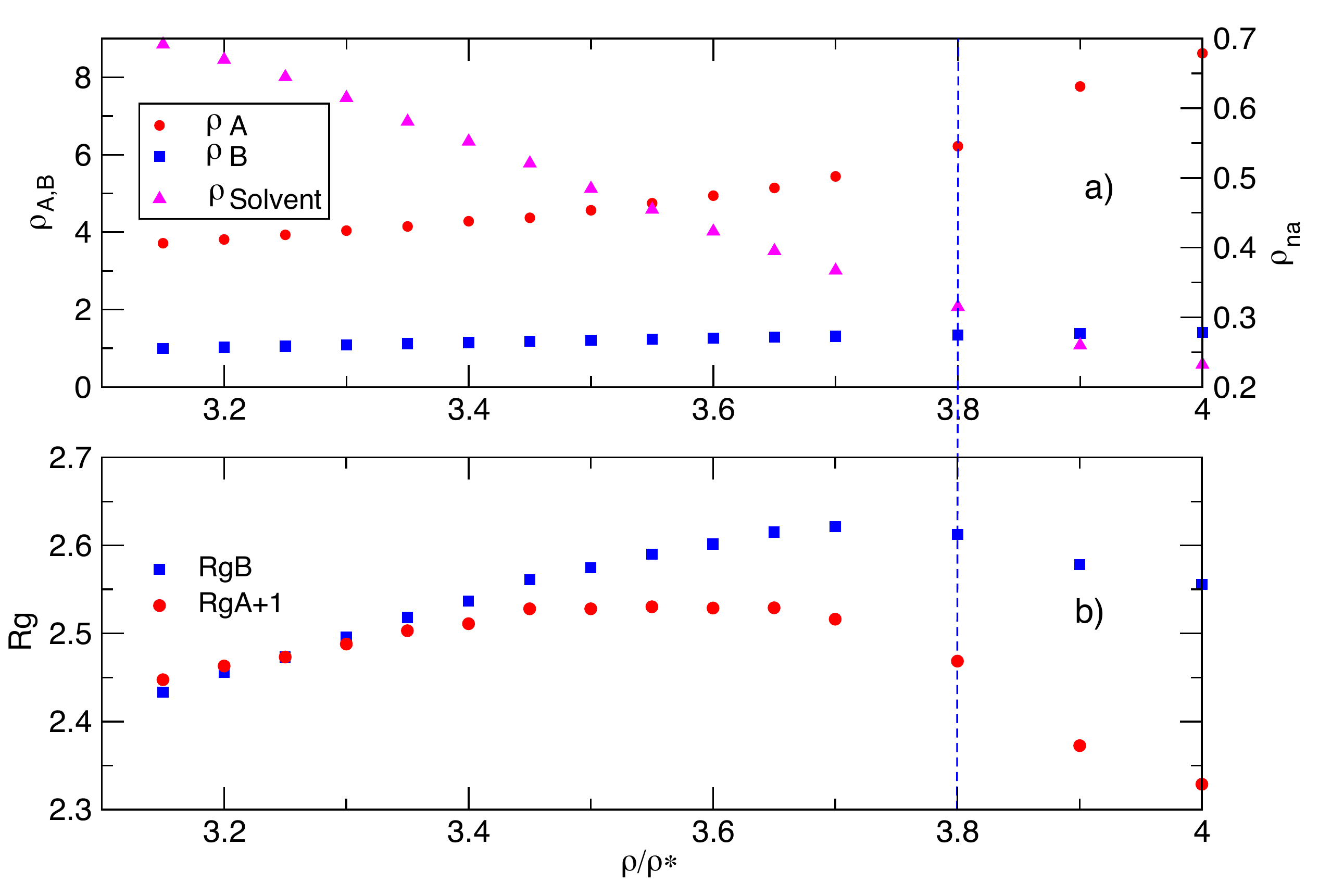}
\caption{ Panel a) Densities of the core $\rho_A$ the corona $\rho_B$ (left axis), and of the dispersed dumbbells  $\rho_{na}$  (right axis) as function of the reduced density; Panel b) Micellar Core and Corona radius as function of the reduced density. For sake of clarity the core radius is shifted by $+1$. The dashed line represent the estimated freezing density.}
\label{fig:dens}
\end{figure}

To further characterize the structure of micelles we report in fig. \ref{fig:dens}(a) the average densities of the micellar core, corona and of the dispersed particles as a function of the reduced density of the system. To compute these densities we have estimated the average volumes of the aggregates, both core and corona, by their radii of gyration. While the corona density remains always $\sim 1$ (in  units of number of molecules over $R_g^3$) the core density is already $\sim 4$ at the CMC and increases up to $\sim 8$ at $\rho/\rho^*=4$. This is an anomaly of our present model due to the absence of any repulsions between A-blocks. The increase of core density up to $\rho/\rho^*\sim 3.5$ is mainly due to the linear increase of the aggregation number (see fig\ref{fig:Nmic}(a)) not compensated by the almost linear increase of the core radius of gyration reported in fig. \ref{fig:dens}(b). Above $\rho/\rho^*\sim 3.5$, the micellar core radius is observed to contract upon increasing the overall density, providing an enhanced increase of core density, even beyond the freezing point above which the aggregation number is almost constant (see fig \ref{fig:Nmic}(a)). The same qualitative behavior is observed for the radius of gyration of the corona although the turning point is at $\rho/\rho^*\simeq 3.7$ and the following decrease is less important.  Finally in fig  \ref{fig:dens}(a) we also report the density of the dispersed molecules computed as the number of dispersed molecules over the volume available to the solvent. The latter has been obtained using the corona radius of gyration to estimate the volume occupied by the aggregates. A linear decrease is observed, meaning that the decrease of available volume partially compensates for the exponential decrease of available molecules observed in fig \ref{fig:Nmic}(b).

\begin{figure}
\center
\includegraphics[scale=0.33]{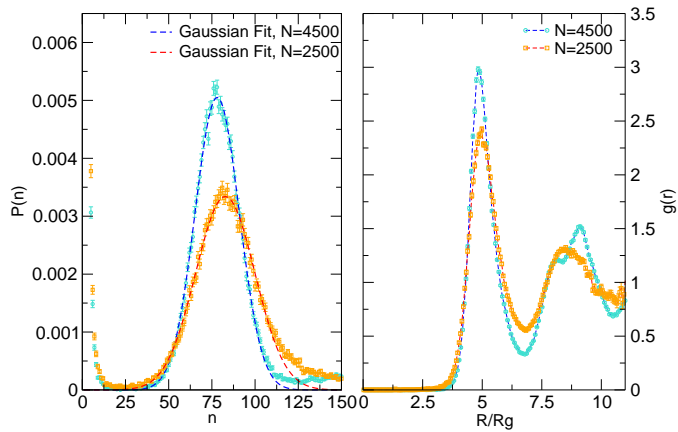}
\caption{Comparisons between the probabilities of finding a cluster of size $n$, $P(n)$ (left panel), and between the micellar radial distribution functions $g(r)$ (right panel) at the same density $\rho/\rho^*=4.$ and different number of dumbbells $N=4500,2500$. }
\label{fig:CLGR}
\end{figure}

In contrast to low density disordered states, the ``experimental" observation is that beyond the freezing point the stationary state depends considerably on the total number of molecules in the system. 
In particular for N in specific ranges (see section \ref{sec:frenergy}) the system is observed to spontaneously order in defective crystalline structures of micelles. As an example, at $\rho/\rho*=4$ a reasonable guess for the number of dumbbells $N$ compatible with the number of micellar sites for a  BCC cluster lattice, allows us to observe the spontaneous emergence of a pattern of peaks in the static structure factor reminiscent of a BCC ordering.
Conversely, for an arbitrary number of dumbbells the system shows a structured S(k) with small peaks and slow structural adjustments, signature of an arrested glassy phase.
Simulations beyond the estimated freezing density for different numbers of molecules reveal a significant change of the $P(n)$ cluster peak, associated with the occurrence of a spontaneously ordered phase. An example is given in figure \ref{fig:CLGR} where we compare $P(n)$ at $\rho/\rho^*=4$ for a BCC phase and a disordered phase. A decrease in the position of the peak and in its width in the ordered phase are evident, accompanied by the appearance of a splitting of the second peak in the radial distribution function between the centres of mass of micelles. This phenomenon is associated with a reduction of  the system polydispersity as counterpart of the change of local and long range order in the crystalline state. This is an aspect of the complex interplay between the long range ordering of the aggregates and the micellar internal structure. 

\section{Free energy calculations}\label{sec:frenergy}
In this section we apply the thermodynamic formalism developed previously to the free energy calculations of cluster crystalline structures of different symmetry such as simple cubic (SC), BCC, FCC, A15 and diamond, and at two densities $\rho/\rho^{*}=4,5$. The preparation of the system in specific crystalline structures has been achieved by applying the same external field described in section \ref{sec:ComputationalAspects} for the $\lambda$ integration procedure but keeping at the same time the intermolecular interactions active. It turns out to be possible for a given density $\rho/\rho*$, crystal symmetry and number of lattice sites $N_{s}$, to find a window for the number of dumbbells $N$ in the system, in which the different clustered phases are dynamically stable during the MC simulation. This is true in all cases but the diamond structure for which, even with the field turned on, the clusters were not found to order according to the symmetry of the external field. The reason is that in the diamond structure the nearest neighbors distance is shorter than in the other structures and would impose the formation of micelles too small to be stable. Even with the external field of reasonable amplitude turned on, the system prefers to occupy only one site over two and form larger aggregates. Upon turning off the field, the SC structure shows a dynamical instability with a rapid melting independently of $N$. The other three structures were dynamically stable and for each of them we have performed several simulation with different $(N,V)$ in the constrained ensemble $(\rho/\rho*,N_{s})$, namely at fixed density, lattice geometry and number of lattice sites. In order to compute the free energy of each thermodynamic constrained state $(N,V,N_{s})$ by the coupling constant integration procedure, see eq.(\ref{eq:free}), further simulations are necessary for different values of the coupling parameter $\lambda$  chosen according to the Gauss-Legendre method. We have used 8 quadrature points and we report the numerical details in appendix I.

In tables \ref{tab:tabII} and \ref{tab:tabIII} we collect the results for the excess free energy, osmotic pressure and chemical potentials, including $\mu_c$, for different structures at the densities $\rho/\rho^*=4$ and 5, respectively. The first observation are the large uncertainties on the determination of $\mu_c$. This is due mainly to the large size of our aggregates since, according to eq. (\ref{eq:equilibrium}), $\mu_c$ is obtained as combination of specific properties, whose uncertainties are quite small, multiplied by the average number of molecules per lattice site $N/N_s$. It is then clear that uncertainties on $\mu_c$ are two orders of magnitude larger than uncertanties on any other intensive thermodynamic property. This fact makes is quite difficult to obtain an accurate determination of the number of particles at which $\mu_c$ vanishes.

At $\rho/\rho^{*}=4$, excess free energies and chemical potentials are largely insensitive to $N/N_{s}$ and to the lattice geometry. Conversely the excess osmotic pressure decreases for increasing $N/N_{s}$ probably due to its dependence on the lattice spacing which also increases with N. Despite the large uncertainties on $\mu_c$, our results allow to identify the free energies of the unconstrained crystals by a simple linear interpolation between positive and negative values of $\mu_c$ (for the A15 structure we assume the point at $\mu_c=0.07\pm0.9$ to be representative of the unconstrained system). The interpolated values for the excess free energies per dumbbell, which allows to determine the most stable structure at the given density, are $f_{BCC}=7.6767(5), f_{A15}=7.6776(6)$ and $f_{FCC}=7.6792(4)$. Therefore our procedure indicates the BCC structure to be the most stable among the three considered, with the A15 structure only slightly above and within the error bars. These are both open structures in contrast to the close packed FCC structure which appears to have an higher free energy. Note that the variation of the Helmholtz free energy among the various structures is of the same order of magnitude as the observed variations with $N/N_s$ within the same structure, so that it would have been impossible to determine the ordering of the structures without determining the unconstrained state at $\mu_c=0$.
  
At  $\rho/\rho^{*}=5$ a similar study for BCC and FCC lattice, as summarized in table \ref{tab:tabIII}, shows a clear dependence of the excess free energy, as well of the osmotic pressure and $\mu_c$, on the number of molecules. However as at lower density, the excess chemical potential is almost insensitive to the number of particles. Even in this case the uncertainties on $\mu_c$ are large but the stability of the crystalline structure for systems with both positive and negative values of $\mu_c$ allows easily to identify the equilibrium unconstrained state by linear interpolation. The extracted values for the excess free energies per dumbbell are $f_{BCC}=9.12(1), f_{FCC}=9.184(3)$ indicating again the BCC structure as more stable for this system.
\begin{figure}[h]
\center
\includegraphics[scale=0.33]{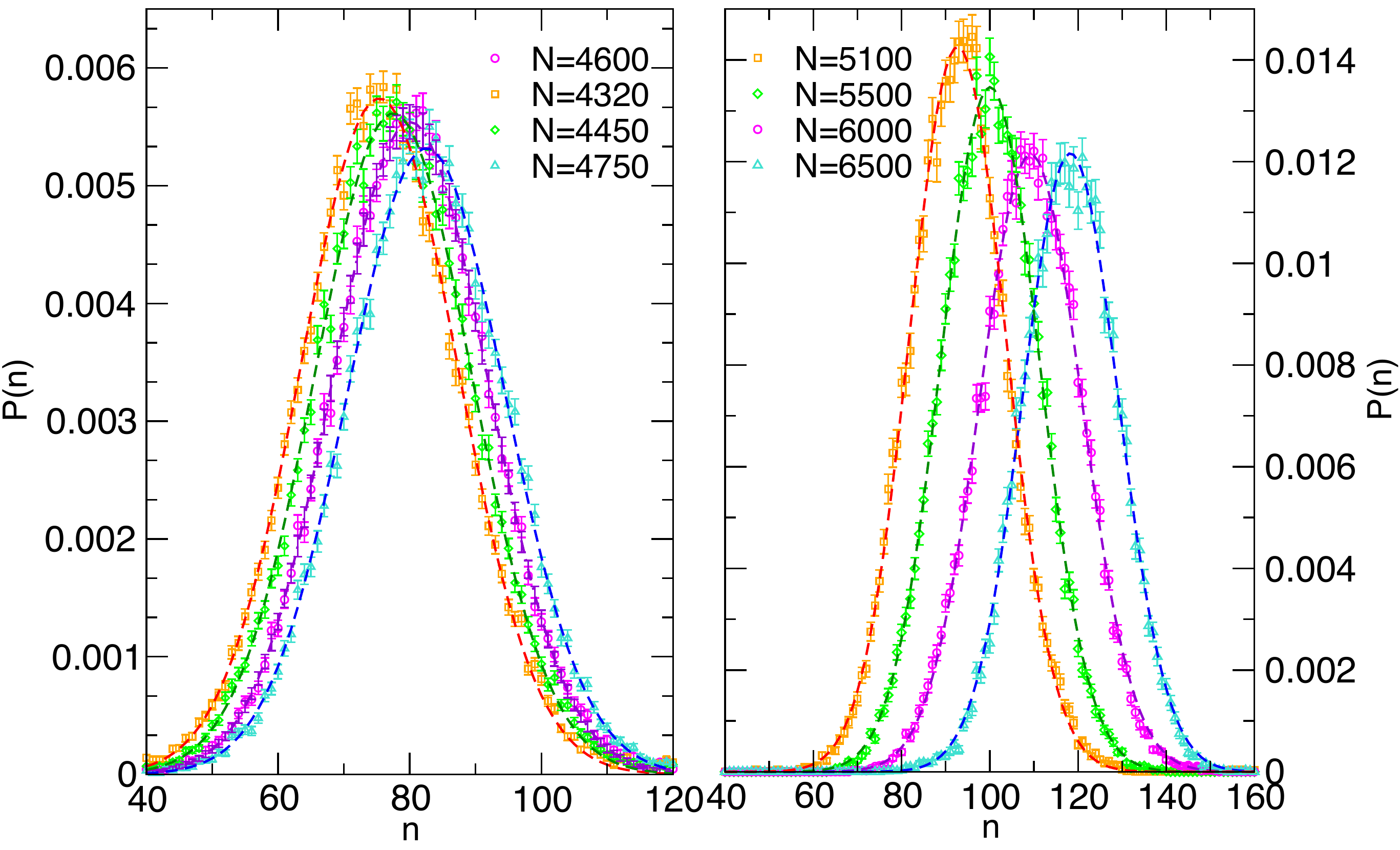}
\caption{BCC structure: comparison of the cluster peaks of $P(n)$ for $\rho/\rho^*=4$ (left) and $\rho/\rho^*=5$ (right) at different numbers of dumbbells. The dashed lines represent gaussian fits to the peaks.}
\label{fig:PN}
\end{figure}

The present procedure to determine the unconstrained equilibrium state, allows to discuss the true equilibrium properties of the micelles in the symmetry broken phase. For the stable BCC phase, the variation of $P(n)$ with the total number of molecules is reported in figure \ref{fig:PN} for the two densities investigated. At both densities a shift of the cluster peak to higher aggregation numbers for increasing $N$ is observed. The shape of the peak is fairly well fitted by a gaussian function with variance independent of $N$. The observed decrease of the amplitude of the peak with $N$ is balanced by an observed increases of the number of dispersed molecules. Numerical results for the mean value and the variance of the fitted gaussian are reported as $n_c$ and its uncertainty in tables \ref{tab:tabII} and \ref{tab:tabIII}, for $\rho/\rho^*=4$ and 5, respectively. The interpolated equilibrium values for the BCC structure are $n_c=81(12)$,$n_c=115(12)$ for $\rho/\rho^*=4$ and $\rho/\rho^*=5$ respectively.

\begin{figure}
\center
\includegraphics[scale=0.33]{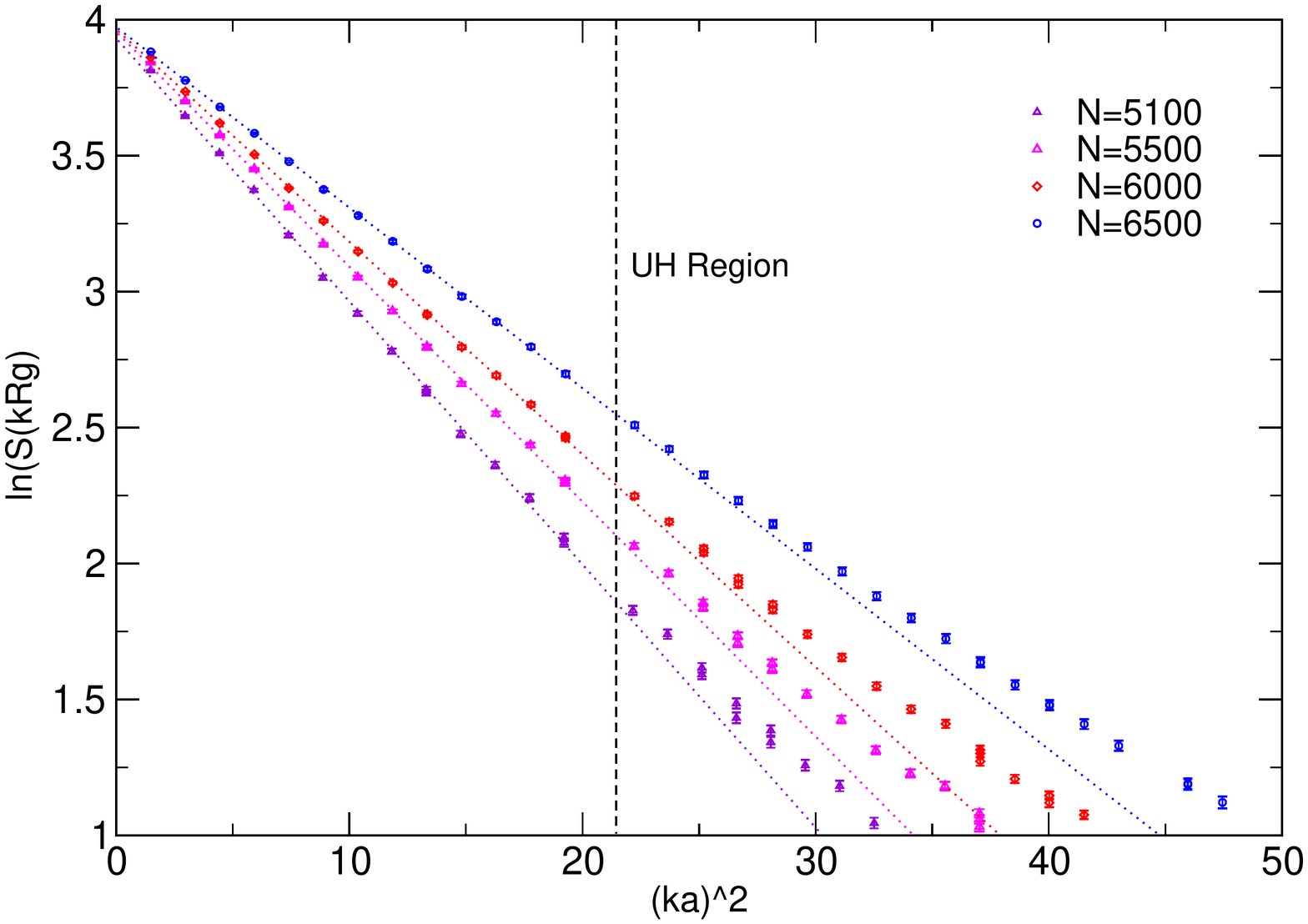}
\caption{Logarithm of the micelle-micelle structure factor $S(k)$  versus the square modulus of the reciprocal vector multiplied by the lattice spacing $a$, for a BCC structure of clusters at ${\rho}/\rho^{*}=5.$, for an increasing number of molecules. The dashed line indicates the onset of anharmonicity.}
\label{fig:DW}
\end{figure}

Tables \ref{tab:tabII} and \ref{tab:tabIII} also report the values of the Lindemann ratio $\mathcal{L}$ and of the average number of vacancies per lattice site, $\langle N_v\rangle/N_s$, for the various structures as a function of $N$. In general for a given structure, both quantities decrease for increasing N meaning that the mechanical stability of the lattice increases with N. However this behavior is limited to a finite range of N values above which the crystal structure is found to spontaneously melt. At $\rho/\rho^*=4$ the Lindemann ratio is significantly  larger than the classical value at melting ($\mathcal{L} \sim 0.15$) and found to be $\mathcal{L}_{fcc} \sim 0.25$ and $\mathcal{L}_{bcc}\simeq\mathcal{L}_{A15}\sim 0.29$ at the unconstrained state.
The equilibrium value 0.29, reminiscent of the value observed for delocalized quantum particles near melting\cite{JonesCeperley06}, suggests that the internal micellar structure plays the role of quantum fluctuations with respect to the vibrations of the centre of the composite particle. Note that we have not computed the free energy of the disordered liquid state at $\rho/\rho^*=4$ but the BCC crystal stability against the liquid phase is supported by the observed spontaneous crystallization of a sample with $N=4500$. According to table \ref{tab:tabII} this system has $\mu_c\approx1$, resulting in a free energy higher than the unconstrained state, and therefore proving the stability of the latter with respect to the disordered phase. 
At $\rho/\rho^*=5$ the same qualitative behavior is observed with smaller values of $\mathcal{L}$'s and $\langle N_v\rangle/N_s$ for all structures meaning that this point is deeper inside the crystal phase. At this density and for the stable BCC phase, we also report in figure \ref{fig:DW} the peaks of the micellar S(k) which allows to directly extract the Debye-Waller factor $W$ \cite{Ashcroft}. We observe a trend of $W$ compatible with the increased stability with N. Note that the equilibrium state does not corresponds to the mechanically most stable state. This is natural since maximal mechanical stability corresponds to the state with no vacancies while the thermodynamic equilibrium state has a non vanishing concentration of vacancies.  

\begin{widetext}
\center
\begin{table}[h!]
\caption{Main thermodynamic and structural properties for different structures and number of dumbbells for $\rho/\rho^*=4.$ and $f=0.6$. $N_{s}$, $n_{c}$, $\mathcal{L}$ and $\langle N_v\rangle /N_s$  represent the number of sites, the mean aggregation number of micelles, the Lindemann ratio and the equilibrium fraction of vacancies respectively.}
 \begin{tabular}{ c c c c c c c c c c c  }
 \hline
 \hline  
  $N_{s}$-Structure & $N$ & $N/N_{s}$ & $n_{c}$ & $F^{ex}/N$ & $\mu^{ex}$ & $\Pi^{ex}/\rho$ & $\mu_{c}$ & $\Delta \mu_{c}$ & $\mathcal{L}$ & $\langle N_v\rangle /N_s$ \\
 \hline
 $32-FCC$ & $2916$ & $91$ & $85(12)$ & $7.6794(4)$  & $13.99(1)$ & $6.3031(8)$ & $-0.7$ & $0.9$ & $0.241(3)$ & $0.049(1)$\\
 $32-FCC$ & $2624$ & $82$ & $76(12)$ & $7.6791(4)$  & $14.04(1)$ & $6.369(1)$ & $0.5$ & $0.8$ & $0.264(3)$ & $0.060(1)$\\
 & & & & & & & \\
$54-BCC$ & $4750$ & $88$ & $83(12)$ & $7.6762(6)$  & $14.01(1)$ & $6.313(1)$ & $-1.8$ & $0.9$  & $0.291(2)$ & $0.062(1)$ \\ 
$54-BCC$ & $4624$ & $85$ & $80(12)$ & $7.6770(5)$  & $14.00(1)$ & $6.332(1)$ & $0.8$ & $0.9$  & $0.295(2)$ & $0.0659(7)$\\
$54-BCC$ & $4450$ & $82$ & $78(12)$ & $7.6782(6)$  & $14.02(1)$ & $6.3556(9)$ & $1.1$ & $0.7$ & $0.306(3)$ & $0.072(1)$\\
$54-BCC$ & $4320$ & $80$ & $75(12)$ & $7.6773(5)$  & $14.02(1)$ & $6.379(1)$ & $2.9$ & $0.8$ & $0.310(2)$ & $0.0759(7)$ \\
& & & & & & & \\
$64-A15$ & $5248$ & $82$ & $77(12)$ & $7.6786(4)$  & $14.03(1)$ & $6.382(1)$ & $2.6$ & $0.8$ & $0.260(2)$ & $0.0422(9)$ \\
$64-A15$ & $5760$ & $90$ & $84(12)$ & $7.6776(4)$  & $14.00(1)$ & $6.323(1)$ & $0.07$ & $0.9$  & $0.290(2)$ & $0.061(1)$ \\
 \hline
 \hline
\end{tabular}
\label{tab:tabII}
 \end{table}

\begin{table}[h!]
\caption{Main thermodynamic and structural properties for different structures and number of dumbbells for $\rho/\rho^*=5.$ and $f=0.6$. The definition of the symbols is the same as in Table \ref{tab:tabII}}
 \begin{tabular}{ c c c c c c c c c  c c }
 \hline
 \hline  
  $N_s$-Structure & $N$ & $N/N_{s}$ & $n_{c}$ & $F^{ex}/N$ & $\mu^{ex}$ & $\Pi^{ex}/\rho$ & $\mu_{c}$ & $\Delta \mu_{c}$ & $\mathcal{L}$ & $\langle N_v\rangle /N_s$  \\
 \hline
 $32-FCC$ & $4300$ & $134$ & $131(11)$ & $9.1881(8)$  & $16.21(1)$ & $6.9836(6)$ & $-5.1$ & $1.3$ & $0.06889(2)$ & $0.0003(1)$ \\
 $32-FCC$ & $4000$ & $125$ & $122(11)$ & $9.1813(8)$  & $16.20(2)$ & $7.0477(6)$ & $3.6$ & $2.5$ & $0.0747(8)$ & $0.0008(1)$  \\
 $32-FCC$ & $3680$ & $115$ & $112(11)$ & $9.1667(8)$  & $16.26(2)$ & $7.1320(4)$ & $4.4$ & $2.3$ & $0.264(3)$ & $0.060(1)$  \\
 $32-FCC$ & $3500$ & $109$ & $107(11)$ & $9.1703(8)$  & $16.29(2)$ & $7.1842(4)$ & $7.0$ & $1.6$  & $0.291(2)$ & $0.062(1)$\\
 & & & & & & & \\
$54-BCC$ & $6500$ & $120$ & $118(11)$ & $9.0885(9)$  & $16.21(2)$ & $7.0722(4)$ & $-5.9$ & $2.4$ & $0.083(2)$ & $0.0017(2)$  \\
$54-BCC$ & $6000$ & $111$ & $109(11)$ & $9.1807(9)$  & $16.25(2)$ & $7.1026(4)$ & $13.7$ & $2.2$ & $0.0901)$ & $0.0020(2)$ \\
$54-BCC$ & $5500$ & $101$ & $100(11)$ & $9.1895(9)$  & $16.28(2)$ & $7.2273(4)$ & $13.9$ & $2.0$ & $0.103(1)$ & $0.0037(2)$\\
$54-BCC$ & $5100$ & $94$ & $93(11)$ & $9.1985(9)$  & $16.38(2)$ & $7.3566(4)$ & $16.5$ & $1.9$ & $0.124(3)$ & $0.0070(5)$ \\
 \hline
 \hline
\end{tabular}
\label{tab:tabIII}
 \end{table}
 \end{widetext}

\section{Conclusion}\label{sec:Conclusion}

We have reported a quantitative investigation of the relative stability of  self-assembling micellar crystalline phases of  diblock copolymer in semi-dilute solutions. We have considered an athermal ISS model in which a copolymer has a block (A) in theta solvent conditions while the other block (B) in good solvent conditions, and the A-B interaction are mutually-avoiding.  
We have adopted the minimal coarse-grained model for such systems, consisting of mapping a long diblock copolymer onto a soft-core dumbbell. The intra- and inter-molecular interactions are obtained in the infinite dilution limit and then used at finite dilution on the basis of an untested transferability assumption. 
Beyond a CMC, the system of dumbbells is found to self-assemble into spherical clusters which form a liquid-like system of micelles. Upon further increase of density a ODT between the micellar liquid and a micellar crystal is met. Our investigation has aimed to determine the most stable crystalline state among several candidate cubic structures. This program requires the calculation of the free energies for the various candidate structures and has been performed by applying to our system a recently proposed method to compute free energies of multiple-occupancy crystals\cite{Mladek1,Mladek2}. Our study has been limited to a single value of the relative block length ratio $f=M_A/(M_A+M_B)=0.6$ and to only two densities $\rho/\rho^*=4$ and $5$, the former close to the (still undetermined) transition density, the latter more inside the crystal phase. The reason is partially the large computational cost of the present method, but more fundamentally the qualitative character of our minimal coarse-grained model to represent diblock copolymer solutions at intermediate density. Therefore the present study is mainly methodological and has prepared the stage for the application of the free energy technique to the more accurate multi-blob  representation\cite{PierleoniCaponeHansen07,CaponeHansenColuzza10}.

We have performed MC simulations for increasing density in the range $\rho/\rho^*\in[3,5]$ with the aims of (i) characterizing the micellar structure with density in the liquid phase, an aspect which has not been fully reported in previous works \cite{Pierleoni06,Capone09}, and (ii) finding the freezing density at $\rho/\rho^*\approx 3.8$, as signaled by the Hansen-Verlet criterion for the $S(k)$ and by a marked change of behavior in the mean aggregation number. Furthermore at $\rho/\rho^*=4$ and $5$, for several crystalline symmetries, we have performed a number of canonical simulations for increasing $N/N_s$ in order to determine the equilibrium state, at which the site chemical potential $\mu_c$ vanishes. This is the unique equilibrium  symmetry broken state of the system that needs to be known in order to compare different crystal symmetries without any bias imposed by the simulation constraints. We have found that our present system prefers the less packed BCC structure over the more packed FCC ones, with the A15 structure, an open non-Bravais structure which minimize the surface to volume ratio of the Wigner-Size cell\cite{WeairePhelan94,Kamien}, almost degenerate with the BCC one. At $\rho/\rho^*=5$, well inside the crystal phase, the ratio of the corona layer thickness to the core radius $L/R_c$, both estimate from the radii of gyration of core and corona, is $L/R_c=1.3(4)$ in good agreement with the experimental value $L/R_c=1.2$ for poly(stirene-b-isoprene)22-12 with f=0.6 (see table I of \cite{BangLodge08}).  Also in qualitative agreement with the experiments, we find that our system prefers the BCC symmetry over the FCC one. Knowing the true equilibrium state of the system we are in the position to compute structural equilibrium properties such as the Lindemann ratio $\cal{L}$, the average number of vacancies per lattice site $\langle N_v\rangle/N_s$ and the Debye-Waller factor $W$. Near melting we observe a large Lindemann ratio ($\sim 0.29$) reminiscent of the value for low temperature quantum particles\cite{JonesCeperley06}, and a substantial number of vacancies per lattice site ($\sim 6\%$). Both quantities decrease when going deeper inside the crystal phase and we observe a residual number of vacancies per lattice site of $\sim 0.2\%$ at $\rho/\rho^*=5$. 

One limitation of our present approach is related to the "small" size of the systems considered. Indeed we examine systems of small number of micelles (32, 54 and 64) and we can expect a finite size effect on the computed free energies. Such effects will not change the ordering of structures far from the melting transition but could be relevant for an accurate location of the melting line and, because of the small free energy differences observed near melting, to determine the stable structure upon crystallization. A possible route to address this problem would be to invoke a second level of coarse graining in which each micelle is replaced by a point particle and the density dependent pair potential between two micelles is extracted by the radial distribution function between the centre of mass of micelles computed in the present dumbbells coarse-grained representation\cite{Pierleoni06,Molina09}. With this further coarse graining, roughly 100 dumbbells are mapped on to a single particle with an important gain in efficiency and the possibility to study systems of thousands of micelles with negligible finite size effects. Despite a number of approximations involved in this further coarse-graining step, namely the neglect of the polydispersity of micelles and of the exchange of polymers between micelles and with the dispersed phase, this procedure was found to reproduce very accurately the structure of the micelles in the liquid phase and its extension to determine the relative stability of several crystal phases has been attempted\cite{Molina09}. It was found a general preference for the A15 structure followed by the FCC structure, while the BCC structure was found to be only marginally stable. However the micelle-micelle potential can be extracted from the structure, only in the continuous symmetry phase and its determination at solid densities was limited by the possibility to prevent the system from crystallizing. We have seen here however that the structure of micelles changes significantly upon crystallization at fixed density, which would imply that the effective micelle-micelle potential would depend not only on the density but also on the symmetry of the state. Unfortunately a procedure to extract a pair interaction from structures with broken symmetries has still to be devised and would require the extension of integral equation theory to the crystalline phase. We believe this limitation in obtaining the pair potentials is the main reason for the disagreement with our present results. 

Very recently the extension of the multi-blob approach\cite{PierleoniCaponeHansen07} to diblock copolymers has been proposed and applied to study the phase diagram of the ISS model for various values of $f$ and at various densities\cite{CaponeHansenColuzza10}. In the crystalline phases, the same dynamical stability in a limited range of $N/N_s$ as observed by us, has been reported and the Helmholtz free energy per polymer was found to be independent on $N/N_s$. On the basis of this independence, the authors concluded that the system is not able to select an optimal cluster size and therefore an optimal lattice spacing. However no evaluation of  the unconstrained equilibrium state at $\mu_c=0$ has been attempted. From our present results we see that $F/N$ and $\mu$ are fairly insensible to $N/N_s$. However this is not proving the inability of the system to select a specific aggregation number and lattice spacing, but it is signaling the presence of the extra constraint imposed by the fixed number of lattice sites in the simulation box. We believe our present analysis, when applied to the multiblob representation, will locate univocally the unconstrained equlibrium state and its properties.

It is well know that temperature is a relevant variable in determining the phase diagram of experimental diblock copolymer solutions. Changing temperature the selectivity of the solvent to the two blocks of the copolymer can be continuously varied, giving rise to thermotropic transitions and to a rich phase diagram. Our present model, being athermal, is not capable to reproduce this phenomenology. The extension of the present coarse-graining strategy to the case of homopolymers in solvent of varying quality has already appeared\cite{Krakoviack}. At variance with the case of chains in good solvent, it was observed that the density dependence of effective potentials between polymers is quite significant and becomes increasingly important when approaching the theta point. This would imply that a minimal dumbbell model of diblock copolymer solutions with zero density effective potentials is a poor representation of the system. However, the temperature effect could be included in the present coarse-graining strategy by adopting the multiblob representation based on zero density potentials and varying the level of coarse-graining according to the chosen temperature. Work along this direction is in progress.

Finally we briefly discuss the relation between the present coarse-graining strategy and the Self-Consisted Mean Field Theory (SCMFT) based methods \cite{Grason,Zhulina05,Suo}. The present simulation approach, although based on coarse-grained models, remains computationally much heavier than SCMFT. Moreover determining free energies requires special methods as shown in the present work, and inferring accurate phase boundaries requires a method to estimate finite size effects still to be developed. Therefore at present SCMFT methods are superior in their ability to reproduce the entire phase diagram with limited effort. Nevertheless, our multiscale strategy to go from the full monomer system to the system of many micelles in a consistent and controlled way provides detailed information on new phenomena such as the interplay between the internal structure of the micelles and their order which has not been addressed previously. This interplay can determine a dependence of the effective micelle-micelle interaction on the ordering of the system which can modify the predictions of the phase diagram based on the effective micelle representation\cite{Grason,Molina09}.

\section*{ACKNOWLEDGEMENTS}
We have the pleasure to thank D. Frenkel, J.-P. Hansen and B. Mladek for useful discussions.

\section{Appendix I}\label{sec:appendixI}
In this appendix we describe the procedure to compute the free energy of the reference system, i.e. a system of non interacting dumbbells in the lattice field provided by eq.(\ref{eq:effectivePhi}). The partition function of this reference system can be expressed as an ensemble average. To this end it is convenient to write the single molecule hamiltonian in terms of molecular coordinates and momenta $(r,R),(p,P)$
\beq
H=\sum_{i=1}^{N}\left[p^2_{i}+\frac{P^2_{i}}{4}+\gamma_{AB}(r_{i})+\Phi_{A}(R_{i}-\frac{r_{i}}{2})+\Phi_{B}(R_{i}+\frac{r_{i}}{2})\right]
\eeq
where $(r,p),(R,P)$ are the internal and center of mass position and momentum of the dumbbell defined respectively as:
\beqa
&R&=\frac{r_A+r_B}{2}\nonumber
\\&r&=r_A-r_B\nonumber
\\&P&=2m\dot{R}\nonumber
\\&p&=\frac{m\dot{r}}{2}
\eeqa
with $r_A$ and $r_B$ being the coordinates of the A and B blocks of the generic dumbbells respectively. From now on we set  $m=1$. Tracing out the momenta the partition function reduces to  the product of the ideal gas term and an average over the equilibrium bond distribution:
\beqa
&.& Q_{N}(V,T)= \nonumber
\\ &=&\frac{1}{\Lambda_{Cm}^{3N}\Lambda_{\mu}^{3N}N!}\int dR^{N}dr^{N}e^{-\beta \sum_{i}^{N}[\gamma_{AB}(r_{i})+\Phi_{A}(R_i-r_{i}/2)+\Phi_{B}(R_i+r_i/2)]} \nonumber
\\
                  &=&\frac {q_{intra}^{N}}{\Lambda_{Cm}^{3N}\Lambda_{\mu}^{3N}N!}\left(\int dRdr \frac{e^{-\beta \gamma_{AB}(r)}}{q_{intra}} e^{-\beta[\Phi_{A}(R-r/2)+\Phi_{B}(R+r/2)]}\right)^N \nonumber
\\
	        &=& \frac {q_{intra}^{N}}{\Lambda_{Cm}^{3N}\Lambda_{\mu}^{3N}N!}\left(\int dR\langle e^{-\beta (\Phi_{A}+\Phi_{B})} \rangle_{{intra}}\right)^{N}
\eeqa
where $q_{intra}=\int dr e^{-\beta\gamma_{AB}(r) }$ is the configurational partition function of a single dumbbell, $\Lambda_{Cm}$ and $\Lambda_{\mu}$ are the de Broglie thermal wave lengths of the center of mass and of the internal coordinate respectively, and $\langle \dots \rangle_{intra}$ indicate an average over the bond distribution funtion $\rho_{intra}(r)=e^{-\beta\gamma_{AB}(r) }/q_{intra}$. The reference free energy follows as:
\beqa
F(\lambda=1)/N&=&-k_{B}T\log\left( \frac {q_{intra}}{\Lambda_{Cm}^{3}\Lambda_{\mu}^{3}}\right)-k_{B}T\log\frac{N}{V}-k_{B}T\log\left(\frac{1}{V}\int dR \langle    e^{-\beta(\Phi_{a}+\Phi_{b})} \rangle_{\rho_{intra}} \right) \nonumber
\\
&=&F_{ideal}(N,V,T)/N-k_{B}T\log\left(\frac{1}{V}\int dR \langle    e^{-\beta (\Phi_{a}+\Phi_{b})} \rangle_{\rho_{intra}}\right)
\eeqa
To obtain the excess free energy it is necessary to perform a MC simulation of a single dumbbell in a volume V at temperature T, sampling the bond vector according to $\rho_{intra}(r)$ and averaging along the stochastic trajectory the quantity $e^{-\beta [\Phi_{A}(R-r/2)+\Phi_{B}(R+r/2)]}$. For a faster convergence center of mass moves with amplitude $V^{\frac{1}{3}}/2$ and unitary acceptance has been implemented.
\section{Appendix II}\label{sec:appendixII}
\begin{figure}[h]
\center
\includegraphics[scale=0.30]{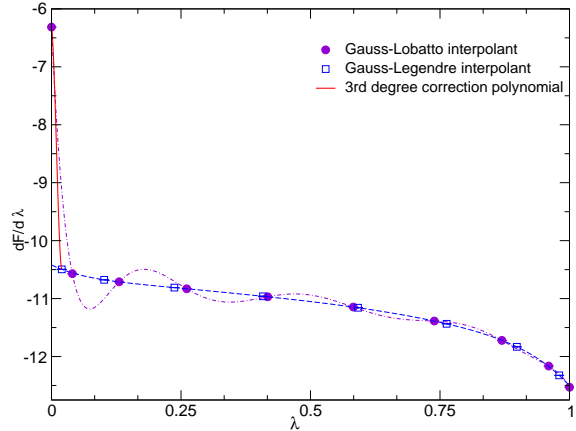}
\caption{Thermodynamic integrand as function of the coupling parameter $\lambda$ for a BCC structure and $\rho/\rho^*=5$. Closed circles indicates the Gauss-Lobatto points, while the open squares the Gauss-Legendre points. The solid, dashed and dot-dashed lines correspond to the 3rd order correction at small $\lambda$, to the Gauss-Legendre and to the Gauss-Lobatto interpolants respectively. }
\label{fig:Gauss}
\end{figure}
In this appendix we report details of the application of the Gauss-Legendre method to compute the $\lambda$-integral in eq. (\ref{eq:free}).
At each $\lambda$ value a $N,V,T$ simulation is necessary to compute $\langle \Phi_{ext}-U_{inter}\rangle_{\lambda} $. 
The softness of the potentials allows us the use of standard single particle moves with large amplitude and hence a good sampling of the previous expression at finite $\lambda$. This is no longer true when $\lambda\rightarrow 0$, because the vanishing of the pinning potential in the Boltzmann weight, progressively allows a slow drift of the system with respect to the location of the field resulting in large fluctuations and slow convergence of the term $\langle\Phi_{ext}\rangle_{\lambda}$. To improve the statistics at $\lambda=0$ we have randomly sampled the geometrical center of the field inside the simulation box. For small but finite $\lambda$-s ($\sim O(1/N)$) this procedure is doomed to fail and a Metropolis sampling of the centre of the field would be necessary but much slower. 
In figure \ref{fig:Gauss} we show a typical behavior of the integrand in eq.(\ref{eq:free}). We note a very smooth behavior with $\lambda$ except at small $\lambda$ where a sharp increase is observed due to the progressive vanishing of the pinning potential.
To accurately integrate this behaviour with a reasonable effort we can either apply the Gauss-Legendre scheme or the Gauss-Lobatto scheme\cite{AbramowitzStegun}, the difference being that in the former scheme only interior points to the interval are used, while the latter scheme employes also the extremes of the interval. In order to compare the accuracy of the Gauss-Legendre scheme and the Gauss-Lobatto one with 8 and 10 points respectively, we report in figure \ref{fig:Gauss} the associated interpolating polynomials.
We note that the interpolating Gauss-Legendre polynomial well represent the expected behaviour, except in a small interval below the first point $\lambda_1$ where we expect a sharp increase toward the value at $\lambda=0$ to occur. Conversely the Gauss-Lobatto interpolating polynomial exhibits an oscillating behaviour which is unphysical. For this reason we have adopted the Gauss-Legendre scheme and implemented a modification to approximately correct for the missing area near the origin. We model the unknown integrand function between the origin and the point $\lambda_1$ by a third order polynomial uniquely defined by matching the values if the integrand function and its first derivative at $\lambda=0$ and at $\lambda=1$. The value of the first derivative at the origin can be estimated by a simple cumulant expansion of the integrand
\beq
\label{cumulant}
\langle\Delta U\rangle_{\lambda}=\langle \Phi_{ext}-U_{inter}\rangle_{\lambda} \simeq
\langle \Delta U\rangle_{0} -\lambda~ \langle\left[\Delta U-\langle\Delta U\rangle_0\right]^2\rangle_0
\eeq
while the derivative in $\lambda_1$ can be obtained as the derivative of the Gauss-Legendre interpolating polynomial. Within this scheme, the corrected estimate of the thermodynamic integral is
\beq
\Delta F=\sum_{i=1}^{8}\langle \Delta U \rangle_{\lambda_{i}} \omega(\lambda_{i})-\mathbb{I}_{GL}(0,\lambda_{1})+\mathbb{I}_{pol}(0,\lambda_{1})
\eeq
where the first term is the Gauss-Legendre quadrature formula, the second term is the numerical integral of the Gauss-Legendre interpolating polynomial between $[0,\lambda_{1}]$ and the last term is the integral over the same interval of the new interpolating polynomial. This correction, of order $\approx \lambda_{1}\langle \Delta U \rangle_{0} $, is essential to obtain an unbiased estimate of the free energy and in our present case turns out to be roughly $0.1\%$ of the value of the total free energy, an order of magnitude larger than the statistical error.

\end{document}